\documentclass[conference,letterpaper]{IEEEtran}

\usepackage[utf8]{inputenc} 
\usepackage[T1]{fontenc}
\usepackage{url}
\usepackage{ifthen}
\usepackage{cite}
\usepackage[cmex10]{amsmath}
\usepackage[utf8]{inputenc} 
\usepackage[T1]{fontenc}
\usepackage{url}
\usepackage{balance}
\usepackage{ifthen}
\usepackage{cite}
\usepackage[cmex10]{amsmath} 

\usepackage[utf8]{inputenc}
\usepackage{amsthm,amsfonts,braket,algorithm,graphicx,multirow}
\usepackage{fullpage}
\usepackage{bbold}

\newcommand{\kb}[1]{\mathbf{[#1]}}
\newcommand{\bkm}[2]{\braket{#1|#2}}
\newcommand{\bk}[1]{\braket{#1|#1}}
\newcommand{\kbr}[1]{\ket{#1}\bra{#1}}

\newcommand{\tensor}{\otimes}
\newcommand{\alphab}{\overline{\alpha}}

\newcommand{\numberthis}{\addtocounter{equation}{1}\tag{\theequation}}

\theoremstyle{definition} 
\theoremstyle{definition} 
\newtheorem {theorem} {Theorem}

\title{Finite Key Analysis of the Extended B92 Protocol}

\author{
  \IEEEauthorblockN{Omar Amer}
  \IEEEauthorblockA{Department of Computer Science and Engineering\\
                    University of Connecticut\\ 
                    06269 Storrs, Connecticut\\
                    Email: omar.amer@uconn.edu}
  \and
  \IEEEauthorblockN{Walter O. Krawec}
  \IEEEauthorblockA{Department of Computer Science and Engineering\\
                    University of Connecticut\\ 
                    06269 Storrs, Connecticut\\
                    Email: walter.krawec@uconn.edu}
 }

\date{December 2019}

\begin{document}

\maketitle
\begin{abstract}
	In this paper we derive a key rate expression for the extended version of the B92 quantum key distribution protocol that takes into account, for the first time, the effects of operating with finite resources. With this expression, we conduct an analysis of the protocol in a variety of different noise and key-length settings, and compare to previous bounds on comparable protocols. 
\end{abstract}
\section{Introduction}
Quantum Key Distribution is becoming an increasingly practically driven field of research \cite{QKD-survey}\cite{QKD-survey-new}. As advances in this and other fields make commercial implementations of QKD devices more desirable, it is necessary that more work is done to understand the capabilities and limitations of these protocols in practice, as opposed to under ideal circumstances. The B92 protocol \cite{B92} has been well researched in the asymptotic setting, where it has been shown to be tolerant to up to $6.5\%$ noise in the channel \cite{MatsumotoB92as}. An extended variant of B92 was proposed \cite{ExtB92}, in which, in addition to the two encoding, non-orthogonal states used in B92, Alice and Bob utilize two additional non-encoding, non-orthogonal states to achieve a tighter bound on Eve's information. Analysis of the extended B92 protocol has shown it to be tolerant to up to $11\%$ noise in the asymptotic setting \cite{ExtB92}. In this paper we will present what is, to our knowledge, the first analysis of the key rate for the extended B92 protocol in the finite key setting. 

Our contributions are as follows: first, we conduct an information theoretic security analysis, assuming a collective attack, to evaluate the key rate of extended B92 in the finite key setting. We use our analysis to rigorously evaluate lower bounds on the key rate and noise tolerance of the protocol in a variety of different channel settings. We note that although we evaluate on depolarization channels, the equations we give hold for arbitrary channels. We will discuss general trends in the optimizing choices of protocol parameters, and, finally, we will compare our findings to the performance of other comparable protocols in the finite key setting. 

\subsection{Notation}
Let $A$ be a random variable, we will denote by $H(A)$ the Shannon entropy of $A$. We will use both $H(p)$ and $h(p)$ to refer to the binary entropy function, and they should both be understood to be equal to $H(p,1-p)$.

Given a pure state $\ket{\psi}_A \in \mathcal{H}_A$ we will use both $\kb{\psi}_A$ and $P(\ket{\psi})_A$ to mean $\kbr{\psi}_A$, and if the context is clear we will often drop the subscript. Given a density operator $\rho_{AB}$ we will write $\rho_B$ to mean the state obtained by taking the partial trace over the $A$ system of $\rho_{AB}$. By a \emph{classical quantum} or CQ state, we will mean a quantum state that can be described by some $\rho_{AB} = \sum_{a}p_a\kb{a}\tensor\rho_B^{(a)}$ for an orthonormal basis $\{\ket{a}\}$. 

Given a density operator $\rho_A$ acting on $\mathcal{H}_A$, we will mean by $S(A)_\rho$ the von-Neumann entropy of $\rho_A$, equivalent to $-tr(\rho_Alog\rho_A)$, where here and elsewhere in this paper $log$ is base $2$ unless otherwise stated. We will mean by $S(A|B)_\rho$ the von-Neumann entropy of the $A$ register of $\rho$ conditioned on the $B$ register, where $S(A|B)_\rho = S(AB)_\rho - S(B)_\rho$. Again, if the context is clear, we may drop the subscript.

Later we will evaluate key rates in a number of channel scenarios, all \emph{symmetric channels}, by which we mean that the channel, parameterized by quantum noise level $Q$, can be described by the depolarization channel
\begin{align}
\mathcal{E}_Q(\rho)\mapsto(1-2Q)\rho+QI. \label{eq:symmetric}
\end{align}

In this work we build towards finding a lower bound of the key rate for the extended B92 protocol. To do this, we make use of the key rate equation, Equation \ref{eq:rawkey}, presented in \cite{renner-finite}, which states that in the finite realm, the key rate, $r'$, of a protocol, under collective attacks, can be calculated as below. We note that as we utilize a different sampling method than was used in \cite{renner-finite}, we must utilize a larger confidence interval than was used in \cite{renner-finite}. Our confidence interval, Equation \ref{eq:uncertainty}, is derived from Hoeffding's inequality.
In \cite{renner-finite} it was shown that for a protocol that has run for $N$ rounds, and resulted in $n\leq N$ raw key bits, the key rate $r'$, can be computed to be 
\begin{align}
	r' = \frac{l(n)}{n}= S_\xi(A|E) - \left(leakEC-\Delta\right)/n \label{eq:rawkey},
\end{align}
where
\begin{align}
	S_\xi(A|E)&=\min_{\sigma_{\bar{A}\bar{E}}\in\Gamma} 
	S(\bar{A}|\bar{E})_\sigma\label{eq:uncertainentropy},
\end{align}
with $\Gamma$ consisting of all $\sigma$ which we could expect to induce statistics that differ by no more than $\xi(m_i)$, except with some probability $\epsilon_{PE}>0$, for any of $\{p_i\}_{i=1}^k$ statistics, each gathered over $m_i$ samples, for: 
\begin{align} 
\xi(m) = \sqrt{\frac{ln\left(2/\left(1-\sqrt[\leftroot{-2}\uproot{2}k]{1-\epsilon_{PE}}\right)\right)}{2m}}\label{eq:uncertainty},
\end{align}
where we take $leakEC$ as the number of bits leaked due to error correction of $n $ raw key bits for a given quantum bit error rate; $\Delta =2log_2(1/[\epsilon-\overline{\epsilon}-\epsilon_{EC})]) +7\sqrt{nlog_2(2/(\overline{\epsilon}-\epsilon_{EC}')}$ are bits lost due to finite key effects; $\epsilon,\epsilon_{EC}$ are user parameters that denote the security parameter of the key and the failure probability of error correction respectively; and $\overline{\epsilon},\epsilon_{PE}$, obeying constraints $\epsilon-\epsilon_{EC} > \overline{\epsilon}>\epsilon_{PE}\geq 0$, can be be chosen so as to maximize the key rate. 

To evaluate the von-Neumann entropy in Equation \ref{eq:uncertainentropy}, we will additionally make use of the following theorem:
\begin{theorem}\label{thm:entropy}(From \cite{krawec2016}):
Let $\rho_{AE}$ be a CQ state acting on $\mathcal{H}_A \tensor \mathcal{H}_E$ that can be written as
\begin{align*} 
	\rho_{AE} = \frac{1}{M} \left(\kb{0}_A\tensor\sum_{i=0}^1\kb{g_0^i}_E
	+\kb{1}_A\tensor\sum_{i=0}^1\kb{g_1^i}_E\right) \numberthis\label{eq:densityop}.
\end{align*}
Then 
\begin{align*}
	S(A|E) &\geq \sum_{i=0}^1 \left(\frac{\bk{g_0^i}+\bk{g1^i}}{N}\right)S_i
\end{align*}
\begin{align*}
	S_i &= 
		\begin{cases}
		  S_i = h\left(\frac{\bk{g_0^i}}{\bk{g_0^i}+\bk{g_1^i}}\right) - h(\lambda_i)  &\substack{\bk{g_0^i} > 0,\\ \bk{g_1^i} > 0}\\    
		  S_i = 0 \text{ else }  
		\end{cases} 
\end{align*}
where
\begin{align*}
	\lambda_i &= \frac{1}{2} + \frac{\sqrt{\left(\bk{g_0^i}+\bk{g_1^i}\right)^2 +4Re^2\bkm{g_0^i}{g_1^i}}}{2\left(\bk{g_0^i}+\bk{g_1^i}\right)}
\end{align*}
\end{theorem}
\section{The Protocol and Key-Rate Computation}

The protocol we analyze is actually a simplified version of the Extended B92 protocol that operates as follows. Alice and Bob utilize the bases $Z = \{\ket{0},\ket{1}\}$ and $A = \{\ket{\alpha},\ket{\alphab}\}$ where  $\ket{\alpha} = \alpha\ket{0} + \beta{\ket{1}}$ and $\ket{\alphab} = \beta\ket{0} - \alpha\ket{1}$,  $0 < \alpha < 1$ is a publicly known parameter of the protocol, and $\beta = \sqrt{(1-\alpha^2)}$. On an iteration of the protocol, with probability $P_{enc}$, also a parameter, this round is a \emph{key round}, and Alice randomly prepares and transmits either the state $\ket{0}$ or $\ket{\alpha}$ to Bob. Otherwise, with probability $(1-P_{enc})$ she sends state $\ket{1}$. Bob chooses to measure his received state in the Z or A basis with equal probability. At the end of a round, Alice notifies Bob if the round was a key round, and, if Bob measured either $\ket{\alphab}$ or $\ket{1}$, Bob notifies Alice that the round was conclusive, otherwise that it was inconclusive. On a conclusive key round, Alice's key bit is 0 if she sent $\ket{0}$ and 1 if she sent $\ket{\alpha}$, and Bob's key bit is 0 if he measured $\ket{\alphab}$, and 1 if he measured $\ket{1}$. If a round is not conclusive, or not a key round, the results are used for channel tomography.

Following $N$ rounds of this protocol, Alice and Bob will share a correlated but noisy raw key string of length $n < N$, as well as $m < N-n $ samples that we will show can be used to estimate various channel statics, obtained from rounds that did not contribute to the key. At this point Alice and Bob follow standard post processing procedures, conducting error correction and privacy amplification to distill an $l(n)$ bit secret key \cite{QKD-survey}\cite{QKD-survey-new}. 

In this section we model the state of the system at the end of a key round so that we may find a lower bound on $S(A|E)$, as is necessary to compute the key rate.  To accomplish this, we also discuss how to estimate the parameters of Eve's attack with statistics that are observed during the course of the protocol, as well as how to calculate the confidence interval that must be minimized over for each of those statistics in the finite case.   

\subsection{Bounding the Conditional Entropy}
To bound the quantity $S(A|E)$ we must first compute a density operator for the system at the end of a key round. Because we are considering collective attacks, Eve's attack can be modeled by unitary operator $U$, acting on a qubit and her ancillary space, initialized as $\ket{\chi_E}$, as follows:
\begin{align*}
	U\ket{0,\chi_E} &\mapsto \ket{0,e_0}+\ket{1,e_1},\\
	U\ket{1,\chi_E} &\mapsto \ket{0,e_2}+\ket{1,e_3}.
\end{align*}
For ease of notation we will make explcit the action $U\ket{\alpha,\chi_E} \mapsto \ket{\alpha, f_0} + \ket{\alphab, f_1}$ where
\begin{align*}
	    \ket{f_0} &= \alpha^2\ket{e_0} + \alpha\beta\ket{e_2} + \alpha\beta\ket{e_1} + \beta^2 \ket{e_3}, \numberthis\label{eq:f0}\\
	    \ket{f_1} &= \alpha\beta\ket{e_0} + \beta^2\ket{e_2} - \alpha^2\ket{e_1} - \alpha\beta \ket{e_3}.\numberthis\label{eq:f1}
\end{align*}

As we are interested in the entropy of Alice's key, we condition on this round of the protocol being a key round. In which case Alice begins the protocol by preparing the transit space $\mathcal{H}_T$ as either $\ket{0}$ or $\ket{\alpha}$, sending it into the channel, and storing her key bit in the register $\mathcal{H}_A$. Eve attacks with $U$ acting on $\mathcal{H}_T\tensor\mathcal{H}_E$, resulting in the joint state: 
\begin{align*}
	\rho_{ATE} &= \frac{1}{2} \kb{0}_{A} 
                    \tensor P(\ket{0,e_0} + \ket{1,e_1})_{TE} \\
                &+ \frac{1}{2} \kb{1}_{A} 
                    \tensor P(\ket{\alpha,f_0} + \ket{\overline{\alpha},f_1})_{TE}.
\end{align*}
Bob now chooses to make a measurement of $\mathcal{H}_T$ in either the Z or A basis, each with equal probability. Again conditioning on this round being a key round, he observes either $\ket{1}$ or $\ket{\alphab}$, corresponding to his key register $\mathcal{H}_B$ being set as 1 or 0 respectively. Tracing out the spaces $\mathcal{H}_T$ and $\mathcal{H}_B$ after we condition on a conclusive measurement, we are left with:

\begin{align*}
\rho_{AE} &= \frac{1}{M}\kb{0}_{A} \tensor 
            \left(
                P(\ket{e_1}) + P(\beta\ket{e_0} - \alpha\ket{e_1}) \right)_E\\
        &+ \frac{1}{M} \kb{1}_{A} \tensor
            \left(
            P(\ket{f_1}) + P(\beta\ket{f_0} - \alpha\ket{f_1})
            \right)_E,\\
\end{align*}
where $M$ is a normalization term we will define shortly.
In accordance with Theorem \ref{thm:entropy}, we can then represent this state in the form given in Equation \ref{eq:densityop}, with:
\begin{align*}
	    M &= \sum_{i=0}^1 \bk{g_0^i} + \bk{g_1^i},\numberthis\label{eq:norm}\\ 
	    \ket{g_0^0} &= \ket{e_1},\numberthis\label{eq:g00}\\
	    \ket{g_0^1} &= \beta\ket{e_0} - \alpha\ket{e_1},\numberthis\label{eq:g01}\\
	    \ket{g_1^0} &= \ket{f_1},\numberthis\label{eq:g10}\\
	    \ket{g_1^1} &= \beta\ket{f_0} - \alpha\ket{f_1} = \alpha\ket{e_1} + \beta\ket{e_3}\numberthis\label{eq:g11}.
\end{align*}

\subsection{Parameter Estimation}
With an operator determined it remains to estimate the various inner products of Eve's states as functions of the observable statistics we gather. It is trivial to find the following identities based on Eve's attack operator:
\begin{align*}
\bk{e_0} &= P_{00}, &\bk{e_1} = P_{01},\\
\bk{e_2} &= P_{10}, &\bk{e_3} = P_{11}.
\end{align*}
Where $P_{ij}$ denotes the probability of Bob measuring $\ket{j}$ after Eve's attack, conditioned on Alice sending the state $\ket{i}$. 

Next we consider the information that can be gained by gathering mismatched statistics\cite{krawec2016}\cite{mismatched}\cite{matsumotomismatched}, gathered from rounds in which Alice and Bob chose to prepare and measure states in mismatched bases. For example, by computing the probability $P_{0\alpha}$ we are able to compute the quantity $Re\bkm{e_0}{e_1}$. Indeed, tracing the evolution of the qubit in that case, we find:
\begin{align*}
	\ket{0} \mapsto &\ket{0,e_0}+\ket{1,e_1}\\
			=&\ket{\alpha} \tensor(\alpha\ket{e_0}+\beta\ket{e_1})+\\
			&\ket{\alphab}\tensor(\beta\ket{e_0}+\alpha\ket{e_1})\\
			\implies P_{0\alpha}= &\alpha^2\bk{e_0}+\beta^2\bk{e_1}\\
			&+2\alpha\beta Re\bkm{e_0}{e_1}\\
			\implies Re\bkm{e_0}{e_1} &= \frac{P_{0\alpha} - \alpha^2\bk{e_0} - \beta^2\bk{e_1}}{2\alpha\beta}.\numberthis \label{eq:e01}
\end{align*}
Similarly we can also find:
\begin{align*}
	Re\bkm{e_2}{e_3} &= \frac{P_{1\alpha} - \alpha^2\bk{e_2} - \beta^2\bk{e_3}}{2\alpha\beta},\numberthis \label{eq:e23}\\
	Re\bkm{e_0}{e_2} &= \frac{P_{\alpha0} - \alpha^2\bk{e_0} - \beta^2\bk{e_2}}{2\alpha\beta},\numberthis \label{eq:e02}\\
	Re\bkm{e_1}{e_3} &= \frac{P_{\alpha1} - \alpha^2\bk{e_1} - \beta^2\bk{e_3}}{2\alpha\beta}.\numberthis \label{eq:e13}
\end{align*}
Through much the same method, utilizing the states given in Equations \ref{eq:f0} and \ref{eq:e02}, we are able to find the following identity using $P_{\alpha\alphab}$.
\begin{align*}
	2\alpha^2&\beta^2Re\left(\bkm{e_0}{e_3} + \bkm{e_1}{e_2}\right) =\\
	&\alpha^2\beta^2\left(\bk{e_0}+\bk{e_3}\right) +\\
	& \beta^4\bk{e_2}+\alpha^4\bk{e_1} +\\
	& 2\alpha^3\beta Re\left(\bkm{e_1}{e_3}-\bkm{e_0}{e_1}\right)+\\
	& 2\alpha\beta^3 Re\left(\bkm{e_0}{e_2}-\bkm{e_2}{e_3}\right) - P_{\alpha\alphab}.\numberthis\label{eq:e03+e12}
\end{align*}
With the last of our identities described, we can now apply Theorem \ref{thm:entropy} to find a lower bound on the entropy of Eve's system to be:
\begin{align*}
	S(A|E) &\geq \sum_{i=0}^1 \left(\frac{E_0[i]+E_1[i]}{N}\right)S_i \numberthis\label{eq:entropybound}
\end{align*}
\begin{align*}
	S_i &= 
		\begin{cases}
		  S_i = h\left(\frac{E_0[i]}{E_0[i]+E_1[i]}\right) - h(\lambda_i) & \substack{\text{if }E_0[i] > 0\\ \text{ and } E_1[i] > 0}\\    
		  S_i = 0 \text{ else }  
		\end{cases} 
\end{align*}
where $A[i]$ denotes indexing into any of the ordered sets $A$ given below, and 
\begin{align*}
	\lambda_i &= \frac{1}{2} + \frac{\sqrt{\left(E_0[i]+E_1[i]\right)^2 +4Re^2\Lambda[i]}}{2\left(E_0[i]+E_1[i]\right)},\\
	    E_0 &= \{\bk{g_0^0}, \bk{g_0^1}\} = \{P_{01}, 1- P_{0\alpha}\}\numberthis\label{eq:E0array},\\
	    E_1 &= \{\bk{g_1^0}, \bk{g_1^1}\} = \{P_{\alpha, \overline{\alpha}}, 1-P_{\alpha0}\}\numberthis\label{eq:E1array},\\
	    \Lambda &= \{\bkm{g_0^0}{g_1^0}, \bkm{g_0^1}{g_1^1}\},\\
	    \Lambda[0] &= \alpha\beta Re\left(\bkm{e_0}{e_1} + \bkm{e_1}{e_3}\right)- \alpha^2\bk{e_1} \\&+ \beta^2Re\bkm{e_1}{e_2}\numberthis\label{eq:lambda1},\\
	    \Lambda[1] &= \alpha\beta Re\left(\bkm{e_0}{e_1} +\bkm{e_1}{e_3}\right) - \alpha^2\bk{e_1} \\&+ \beta^2Re\bkm{e_0}{e_3}\numberthis\label{eq:lambda2}.
\end{align*}
We note that all of the inner products above, with the exception of $\bkm{e_1}{e_2}$ in Equation \ref{eq:lambda1}, can be estimated by the statistics gathered in this protocol, either having been made explicit in earlier discussion or, in the case of Equations \ref{eq:E0array} and \ref{eq:E1array}, can be computed to be as we claim by further tracing of the evolution of the state. We can now compute the bound given in \ref{eq:entropybound} by minimizing over the sole free variable, which itself can be bounded by Cauchy-Schwartz as $\bkm{e_1}{e_2} \in \left[-\sqrt{\bk{e_1}\bk{e_2}},\sqrt{\bk{e_1}\bk{e_2}}\right],$
with $\bkm{e_0}{e_3}$ obtained by Equation \ref{eq:e03+e12}.
\
\subsection{Finite Key Effects}

To calculate the key rate in the finite case, we must account for uncertainty in our observed statistics, and consider all possible attacks Eve may have used that induce statistics within the relevant confidence interval, as given by Equation \ref{eq:uncertainty}. Let each statistic $P_{ij}$ have been sampled over $C_{ij}$ samples, then, following the work done in \cite{renner-finite}, we find that to calculate a worst case bound on Eve's information we must further minimize the entropy expression given in Equation \ref{eq:entropybound}, now replacing all observed $P_{ij}$ used in parameter estimation with  
\begin{align*}
\hat{P_{ij}} \in (P_{ij}-\xi(C_{ij}),P_{ij}+\xi(C_{ij})),
\end{align*}
save for $\hat{P_{00}}$, $\hat{P_{11}}$, and $\hat{P_{\alpha1}}$ which we take to be equal to $1-\hat{P_{01}}$, $1-\hat{P_{10}}$, and $1-\hat{P_{\alpha0}}$ respectively. This minimization results in a new worst case bound on Eve's uncertainty, correct with probability $1-\epsilon_{PE}$, which we denote $S_\xi(A|E)$.

With $S_\xi(A|E)$, we can now calculate the finite key-length rate, $r'$ with Equation \ref{eq:rawkey}, with the constraints discussed with Equation \ref{eq:rawkey}, though, for our purposes, are more concerned with evaluating the effective key rate, 
\begin{align}
	r=\frac{r'n}{N},\label{eq:effkey}
\end{align}
rather than the key rate itself, where n is the number of raw key bits.
\section{Evaluation}
With a key rate equation finalized, we now consider the key rates that are realizable at various noise and and signal size scenarios. We consider a symmetric channel, as defined in Equation \ref{eq:symmetric} parameterized on quantum noise level $Q$, though we note the equations we have derived thus far hold for arbitrary channels. We calculate the expected number of samples $C_{ij}$ that contribute to statistic $P_{ij}$, for a given $\alpha$ and $P_{enc}$ over $N$ rounds below:
 \begin{align*}
    C_{01} = C_{\alpha\overline{\alpha}} &= \frac{P_{enc}Q}{4}N, \\
    C_{10} &= \frac{(1-P_{enc})Q}{2}N, \\
    C_{0\alpha} = C_{\alpha0} &= \frac{P_{enc}(Q+(1-2Q)\alpha^2)}{4}N,\\
    C_{1\alpha} &= \frac{(1-P_{enc})(Q+(1-2Q)(1-\alpha^2))}{2}N,\\
    C_{k} &= \frac{P_{enc}(Q+(1-2Q)(1-\alpha^2))}{2}N,
\end{align*}
where we use $C_k$ to denote the number of samples that contribute to the raw key. We also note that in practice these values would be observed, and we utilize these expressions only to calculate what they might be expected to be for the purposes of our evaluation. 

We will conduct our analysis with $leakEC = 1.2h(QBER)$ to account for practical inefficiencies in error correction protocols, where QBER is the error rate of the raw key string, for which we will use a worst case upper bound of:
\begin{align}
QBER \leq \frac{P_{01}+\xi({C_{01}})+ P_{\alpha \alphab}+\xi({C_{\alpha\alphab}})}{p_{acc}}\label{eq:QBER},
\end{align}
where 

\begin{align*}
p_{acc} &= P_{01}+\xi({C_{01}}) + P_{\alpha \alphab}+\xi({C_{\alpha\alphab}})\\
& + 2 - (P_{0\alpha}+\xi({C_{0\alpha}})+ P_{\alpha 1}+\xi({C_{\alpha1}})).
\end{align*}
Further, in our analysis, we fix the user parameters $\epsilon = 1\times10^{-9}$ and $\epsilon_{EC} = 1\times10^{-10}$. Additionally, we fix the optimizable parameters $\bar{\epsilon} = 8\times10^{-10}$ and $\epsilon_{PE} = 7\times10^{-10}$. Finally, we numerically optimized over $\alpha$ and $P_{enc}$ in each case to find an optimal effective key rate in various noise level and signal number. 
\begin{figure}
  \includegraphics[width=\linewidth]{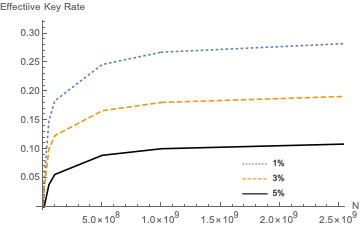}
  \caption{This figure depicts the effective key rate, optimized over $\alpha$ and $P_{enc}$ for quantum noise levels $Q \in \{.01,.03,.05\}$ and evaluated at $N=1\times10^{n}$ and $N=5\times10^{n}$ for $n \in \{6,7,8,9\}$. }
  \label{fig:NvsK}
\end{figure}
\begin{figure}
  \includegraphics[width=\linewidth]{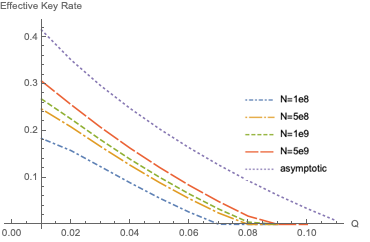}
  \caption{This figure depicts the effective key rate, optimized over $\alpha$ and $P_{enc}$ for various $N$, as well as the asymptotic case (the top line), as noise in the channel increases.}
  \label{fig:QvsK}
\end{figure}
\begin{figure}
  \includegraphics[width=\linewidth]{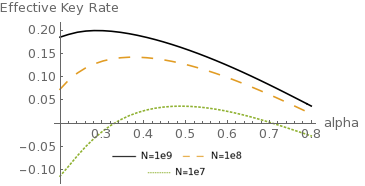}
  \caption{This chart shows effective key rate as $\alpha$ varies for a fixed $P_{enc} = .8$ at noise level $Q = .02$ for various $N$. We found that while in the asymptotic case, the optimal $\alpha$ approaches 0 as shown in \cite{ExtB92}, while in the finite case there is an advantage in optimizing over $\alpha$ (and indeed over $P_{enc}$) in each scenario}.
  \label{fig:alphavsK}
\end{figure}
In Figure \ref{fig:NvsK} we show the optimal effective key rate at various noise levels, increasing with $N$, appearing to numerically approach the asymptotic bound (not shown) at each noise level. In Figure \ref{fig:QvsK}, we show the effective key rate for various $N$ as noise increases, where we can see an increasing effective key rate and noise tolerance as $N$ increases, again approaching the asymptotic bound. 

In our analysis, we observed that the values of $\alpha$ and $P_{enc}$ that led to the optimal key rate (Equation \ref{eq:rawkey}) did not necessarily result in the optimal effective key rate. Additionally we observed that, as $N$ increased, the optimal $\alpha$ decreased while the optimal $P_{enc}$ increased, approaching the asymptotic optimal values of $0$ and $1$ respectively\cite{ExtB92}, as one might expect. Further, we found that for a given $P_{enc}$, the key rate varied with $\alpha$ as shown by the curves in Figure \ref{fig:alphavsK}, reaching no more than one positive maximum.

As this is the first analysis of extended B92 in the finite setting, we instead compare our results to the performance of standard B92 and BB84 in finite settings. As one might expect, our analysis shows that the extended variant of B92, which utilizes additional quantum states to better bound $S(A|E)$, results in higher noise tolerance and effective key rates in the finite setting than can be obtained with standard B92. Indeed, in \cite{matsumoto}, a recent analysis showed that with $10^8$ signals, standard B92 achieves a positive key rate up to at least $6.4\%$ noise while our analysis shows that extended B92 has a noise tolerance of at least $7\%$. Conversely, while the work done in \cite{renner-finite} shows that at $5\%$ noise BB84 can achieve positive key rates with as few as $10^5$ signals, we do not achieve positive rates at that noise until $10^8$ signals.

\section{Closing Remarks}
In this work we have conducted, for the first time, a rigorous, information theoretic finite key-length analysis of a simplified version of the extended B92 protocol. We have bounded the key rate, under collective attacks, for arbitrary channels, and evaluated that bound in various noise scenarios under a symmetric channel. We have shown that the key rate can be improved by optimizing over $P_{enc}$ and $\alpha$, and noted that the optimal choices for those parameters obey interesting trends.

Future areas of interest in this area include refactoring this analysis to utilize a single POVM for gathering statistics, so as to obtain a tighter confidence interval in Equation \ref{eq:uncertainty} as was done in \cite{renner-finite}. Further, it may be possible to achieve higher key rates with a tighter bound on QBER than was given in Equation \ref{eq:QBER}. An analysis of achievable key rates and optimal choices under arbitrary channels may also lead to interesting results, as would an investigation of where optimal values for $\bar{\epsilon}$ and $\epsilon_{EC}$ lie, which we held fixed in our optimization. 
\balance
\bibliographystyle{IEEEtran}
\bibliography{finite.bib}


\end{document}